\documentclass[%
superscriptaddress,
 amsmath,amssymb,
 aps,
 prb,
twocolumn,
]{revtex4-1}

\usepackage{graphicx}
\usepackage{dcolumn}
\usepackage{bm}
\usepackage{hyperref}
\usepackage[mathlines]{lineno}
\usepackage{color}
\usepackage{MnSymbol}

\begin{document}

\title{Self-Templating Assembly of Soft Microparticles into Complex Tessellations}

\author{Fabio Grillo}
\thanks{These authors contributed equally to this work}
\author{Miguel-Angel Fernandez-Rodriguez}\thanks{These authors contributed equally to this work}
\author{Maria-Nefeli Antonopoulou}\thanks{These authors contributed equally to this work}
\author{Dominic Gerber}
\author{Lucio Isa}
\affiliation{Laboratory for Soft Materials and Interfaces, Department of Materials, ETH Zurich, Zurich, 8093, Switzerland}
\email{fabio.grillo@mat.ethz.ch}
\email{ma.fernandez@mat.ethz.ch}
\email{lucio.isa@mat.ethz.ch}


\maketitle

\textbf{\small{Self-assembled monolayers of microparticles encoding Archimedean and non-regular tessellations promise unprecedented structure-property relationships for a wide spectrum of applications in fields ranging from optoelectronics to surface technology\cite{Granick2011,RN5086,Bohlein_2012,RN5087,RN5088,RN5091}. Yet, despite numerous computational studies predicting the emergence of exotic structures from simple interparticle interactions\cite{RN3691,RN5085,RN5092,RN3702,RN3707,Likos_2002,RN3685}, the experimental realization of non-hexagonal patterns remains challenging\cite{RN3708,Rey_JACS_2017,Hummel_2019,Volk_2019}. Not only kinetic limitations often hinder structural relaxation, but also programming the inteparticle interactions during assembly, and hence the target structure, remains an elusive task. Here, we demonstrate how a single type of soft polymeric microparticle (microgels) can be assembled into a wide array of complex structures as a result of simple pairwise interactions. We first let microgels self-assemble at a water-oil interface into a hexagonally packed monolayer, which we then compress to varying degrees and deposit onto a solid substrate. By repeating this process twice, we find that the resultant structure is not the mere stacking of two hexagonal patterns. The first monolayer retains its hexagonal structure and acts as a template into which the particles of the second monolayer rearrange to occupy interstitial positions. The frustration between the two lattices generates new symmetries. By simply varying the packing fraction of the two monolayers, we obtain not only low-coordination structures such as rectangular and honeycomb lattices, but also rhomboidal, hexagonal, and herringbone superlattices which display non-regular tessellations. Molecular dynamics simulations show that these structures are thermodynamically stable and develop from short-ranged repulsive interactions, making them easy to predict, and thus opening new avenues to the rational design of complex patterns.}}

\begin{figure*}
    \centering
    \includegraphics[width=0.8\textwidth]{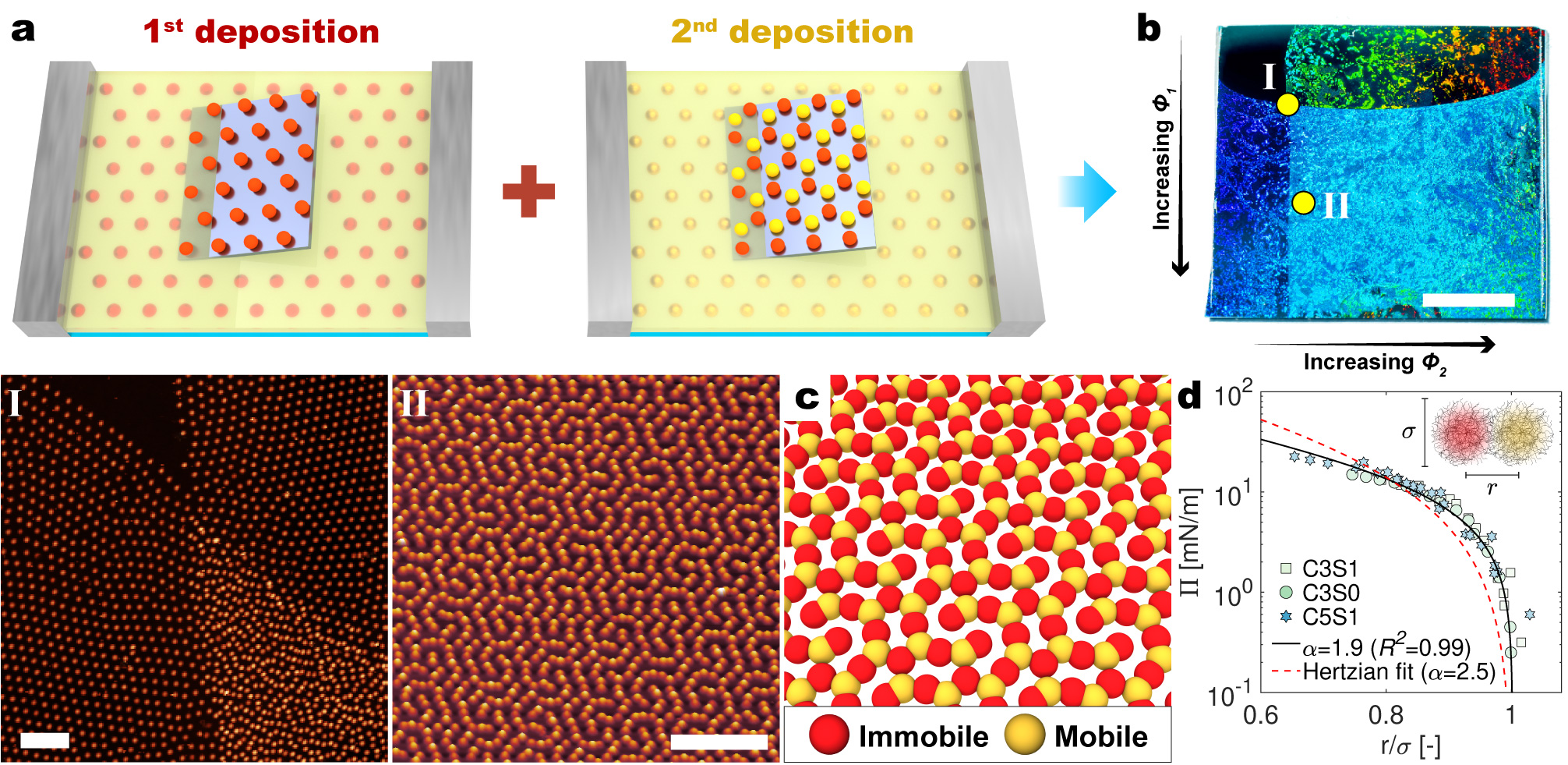}
    \caption{\textbf{Working principles of self-templating assembly of microgels into 2D complex patterns. }\textbf{(a)} Schematic of the double deposition process. \textbf{(b)} Photograph of a silicon substrate after orthogonal deposition of two monolayers (scale bar 0.5 cm). Panels I) and II) show AFM images taken from regions where the two monolayers start overlapping and where $\phi_1>\phi_2$, respectively (scale bars: 5 $\mu$m). \textbf{(c)} Snapshot of the ground-state structure obtained via molecular dynamics simulations for $\phi_1=1.35$, $\phi_2=0.9$ and $\alpha=1.9$ (the particles are drawn with a diameter of 0.6$\sigma$, which  corresponds approximately to the size of the microgels' cores visible in the AFM images). \textbf{(d)} Surface pressure ($\Pi$) vs normalized interparticle distance ($r/\sigma$) from the compression of individual monolayers, fitted to Equation \ref{eq:comp}.}
    \label{fig:Fig1}
\end{figure*}

We synthesized Poly(N-isopropylacrylamide) (PNIPAM) microgels by one-pot precipitation polymerization. The synthesis was tuned to obtain monodisperse microgels of sizes ranging from 600 to 900 nm in MilliQ water (see Methods). Such microgels develop a marked core-corona architecture upon adsorption and confinement at a water-oil interface\cite{Destribats_2011,Geisel_2012, Camerin_2019}, as the portion of the microgels in contact with the interface stretches out radially\cite{Style_2015}. For our particles, the cross-sectional diameter at the interface $\sigma$ is  $\sim1.5-1.8$ times their diameter in bulk aqueous suspensions (see Table \ref{table:mgel_size}). 

The microgels were first assembled into hexagonally packed monolayers of varying packing fractions and then immobilized onto silicon wafers via a modified Langmuir-Blodgett deposition technique. In brief, a given amount of microgels is injected at a water-hexane interface where they these self-assemble into a hexagonally packed monolayer. Such monolayer is then gradually compressed to increasing extents while being transferred onto a silicon wafer, which is lifted through the water-hexane interface (for more details see the Methods section). In this manner, monolayers that are subjected to different surface pressures $\Pi$, and thus possessing different packing fractions, are seamlessly transferred to different locations along one direction of the same substrate\cite{Geisel_SoftMatter_2014, Rey_SoftMatter_2016} (see Fig. S1). Atomic force microscopy (AFM) images of dried microgels after deposition are used to extract the monolayer structure. Only the microgel cores (of size $\simeq 0.6\sigma$) are visible in the images, because the coronas have a thickness of just few nm (see Fig. \ref{fig:Fig1} I-II).

Using an analogous protocol, we compressed and deposited a second monolayer onto the same substrate to combine monolayers of different packing fractions (Fig. \ref{fig:Fig1}a). In particular, we realized orthogonal gradients of packing fraction $\phi_i$ along the two axes of a substrate (Fig. \ref{fig:Fig1}b) by rotating the latter by 90$^{\circ}$ in between depositions. Here, $\phi_i$ is defined as $\frac{n_i}{A}\frac{\pi\sigma^2}{4}$, where $n_i/A$ is the number of microgels per unit area in the first or the second monolayer, and $\sigma$ is the diameter of an isolated microgel at the interface. In addition, by lifting the substrate across the water-hexane interface prior to injecting the microgels, we obtained two bands on the substrate with only one monolayer from each deposition, which we used to estimate the position-dependent $\phi_i$ across the whole substrate (see Fig. S2).

We find that such a double deposition process leads to non-hexagonal two-dimensional patterns that would not otherwise emerge during the compression/deposition of an individual monolayer (see Fig. S3). AFM images reveal that the microgels of the second monolayer are co-planar with the ones of the first monolayer, instead of undergoing out-of-plane stacking (see  Fig. S3 and Figure \ref{fig:Fig1}II). Moreover, while the microgels of the first monolayer retain their hexagonal arrangement, due to the strong adhesion to the underlying substrate \cite{Fernandez-Rodriguez_Nanosacle_2018}, the microgels of the second monolayer can re-arrange and break their hexagonal ordering. As we will show later, the degree of mismatch between $\phi_1$ and $\phi_2$, as well as the total packing fraction $\phi_1+\phi_2$, regulates the formation of a wide spectrum of non-hexagonal patterns. 
We rationalize the formation of such non-hexagonal structures by hypothesizing that the sequential deposition protocol is equivalent to the annealing of a colloidal monolayer comprising two populations of particles that are identical except for the fact that one population is fixed into a hexagonal lattice (see Fig. \ref{fig:Fig1}c). The immobile fraction therefore acts as a template by defining an effective potential energy landscape that frustrates the ordering of the mobile particles, thereby dictating their spatial organization. In particular, we assume that the two populations interact via repulsive short-range pairwise interactions within the same plane. In this framework, the ground-state configuration depends solely on the packing fraction of both monolayers ($\phi_1$ and $\phi_2$), and on the functional form of the pair potential. 

Currently, a model that predicts the pair potential of microgels confined at a fluid interface does not exist. Nonetheless, microgels in bulk suspensions are often described as Hertzian spheres in the limit of small compressions \cite{ZaccaReview2019}. The Hertzian potential then becomes inaccurate for large overlaps between particles, underestimating the repulsion force~\cite{RN3699,RN5078}. We therefore choose to model the repulsive interactions through the generalized Hertzian potential (GHP)~\cite{RN3707,RN3691}:
\begin{equation}
U(r)=\frac{\epsilon}{\alpha}\left(1-\frac{r}{\sigma}\right)^{\alpha}\Theta\left(1-\frac{r}{\sigma}\right)
\label{eq:GHP}
\end{equation}
where $\epsilon$ is the energy scale, $\Theta$ is the Heaviside step function and $\alpha$ is a power-law exponent (equal to 5/2 for the Hertzian case). In particular, we adjust $\alpha$ to capture non-Hertzian behaviours, while still assuming monotonic and short-range repulsive interactions. The value of $\alpha$ for our system is extracted by quantitatively comparing the observed experimental structures with minimum energy configurations ($T=0$) estimated by molecular dynamics simulations (more details in the Methods section). We find that the GHP captures both the mechanical behaviour of individual hexagonally packed monolayers adsorbed at a water-hexane interface and the emergence of the complex structures arising from the assembly of two monolayers. 

\begin{figure*}
    \centering
    \includegraphics[width=0.8\textwidth]{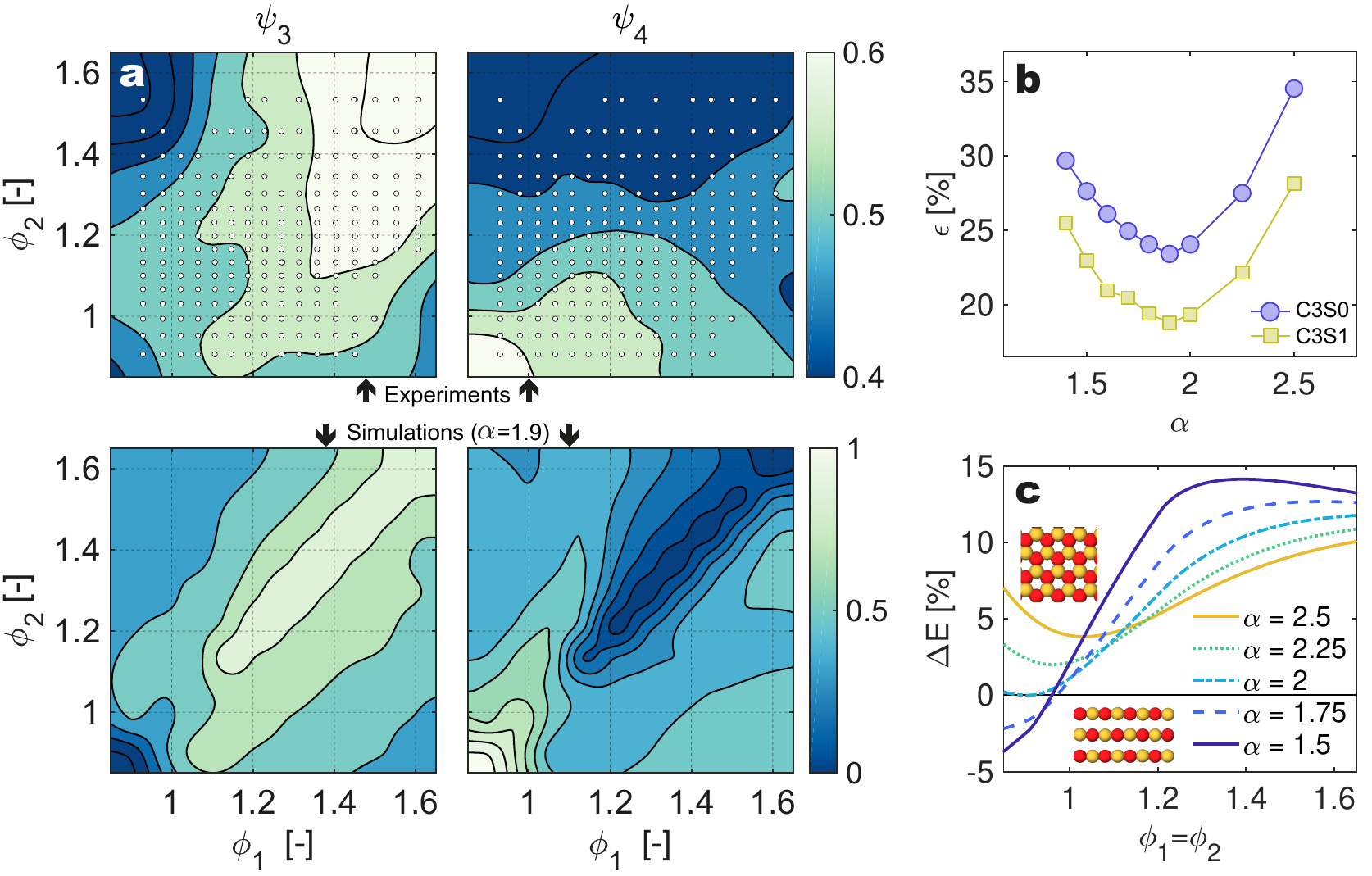}
    \caption{\textbf{Matching experimental and numerical structures.}\textbf{(a)} $\psi_3$ and $\psi_4$ of experimental and ground-state structures as a function of $\phi_1$ and $\phi_2$. The experimental structures were obtained with 3CS1 microgels (analogous trends are found for 3CS0 microgels, see Fig. S5), while the ground-state structures were obtained for $\alpha=1.9$. \textbf{(b)} Overall percentage error $\epsilon$ (see Methods) between simulated and experimental $\psi_k$ as a function of $\alpha$. \textbf{(c)} Percentage difference between the energy density of rectangular and honeycomb lattices as a function of $\phi_1=\phi_2$, for different values of $\alpha$.}
    \label{fig:Fig2}
\end{figure*}

We first tested the validity of the GHP by correlating the surface pressure measured in the Langmuir trough with the inter-particle distance $r$ via a simple analytical expression. Increasing the surface pressure from $\sim$0.2 to $\sim$20 mN/m causes a continuous decrease of the lattice constant of a hexagonally packed monolayer\cite{Geisel_SoftMatter_2014, Rey_SoftMatter_2016}, as $\phi$ increases from $\sim$0.9 to $\sim$1.6 (see Fig. S3). Given that the potential energy and the area of a unit cell of hexagonally packed disks are $E_{hex}=3U(r)$ and $A_{hex}=r^2\sqrt{3}/2$\cite{RN3690}, respectively, we can write the surface pressure $\Pi$ as:
\begin{equation}
    \Pi|_{T=0}=-\frac{\partial E_{hex}}{\partial A_{hex}}=-\frac{\partial U}{\partial r}\frac{\sqrt{3}}{r}=\frac{\epsilon\sqrt{3}}{\alpha\sigma r}\left(1-\frac{r}{\sigma}\right)^{\alpha-1}\Theta\left(1-\frac{r}{\sigma}\right)
    \label{eq:comp}
\end{equation}
neglecting Brownian contributions to the pressure ($\epsilon >> k_{B}T$). Equation \ref{eq:comp} gives an excellent description of the $\Pi-r$ compression curves measured for the three different type of microgels studied, where the fitted values of $\sigma$ are in agreement with the diameter of the microgels at the interface as measured via AFM (see Table \ref{table:mgel_size}). Normalizing $r$ by the respective $\sigma$ causes all the experimental data to collapse onto a single master curve, which is best described by $\alpha= 1.8\pm 0.1$, indicating a significant departure from the Hertzian model ($\alpha=2.5$) (see Fig. \ref{fig:Fig1}d). 

Determining that $\alpha<2$ is particularly significant, as $\alpha=2$ demarcates a qualitative change in the shape of the GHP potential, and thus in the topology of the phase diagram with respect to the structural variety of the ground states \cite{RN3707,RN3694,Likos_Lowen_PRL}. This is because the repulsive force $F(r)=-\partial U/\partial r$ is convex ($\partial^2 F/\partial^2 r>0$) for $\alpha>2$, and concave ($\partial^2 F/\partial^2 r<0$) for $\alpha<2$ (see  Fig. S4). In other words, for $\alpha<2$, the force experienced by two approaching particles increases more rapidly for small overlaps $r\sim\sigma$ than it does for large overlaps $r\sim 0$. This feature can translate into the stabilization of asymmetric and low-coordinated structures such as rectangular lattices that would otherwise be inaccessible for $\alpha>2$\cite{RN3707,Likos_2002,RN5080}. 

\begin{figure*}
    \centering
    \includegraphics[width=0.8\textwidth]{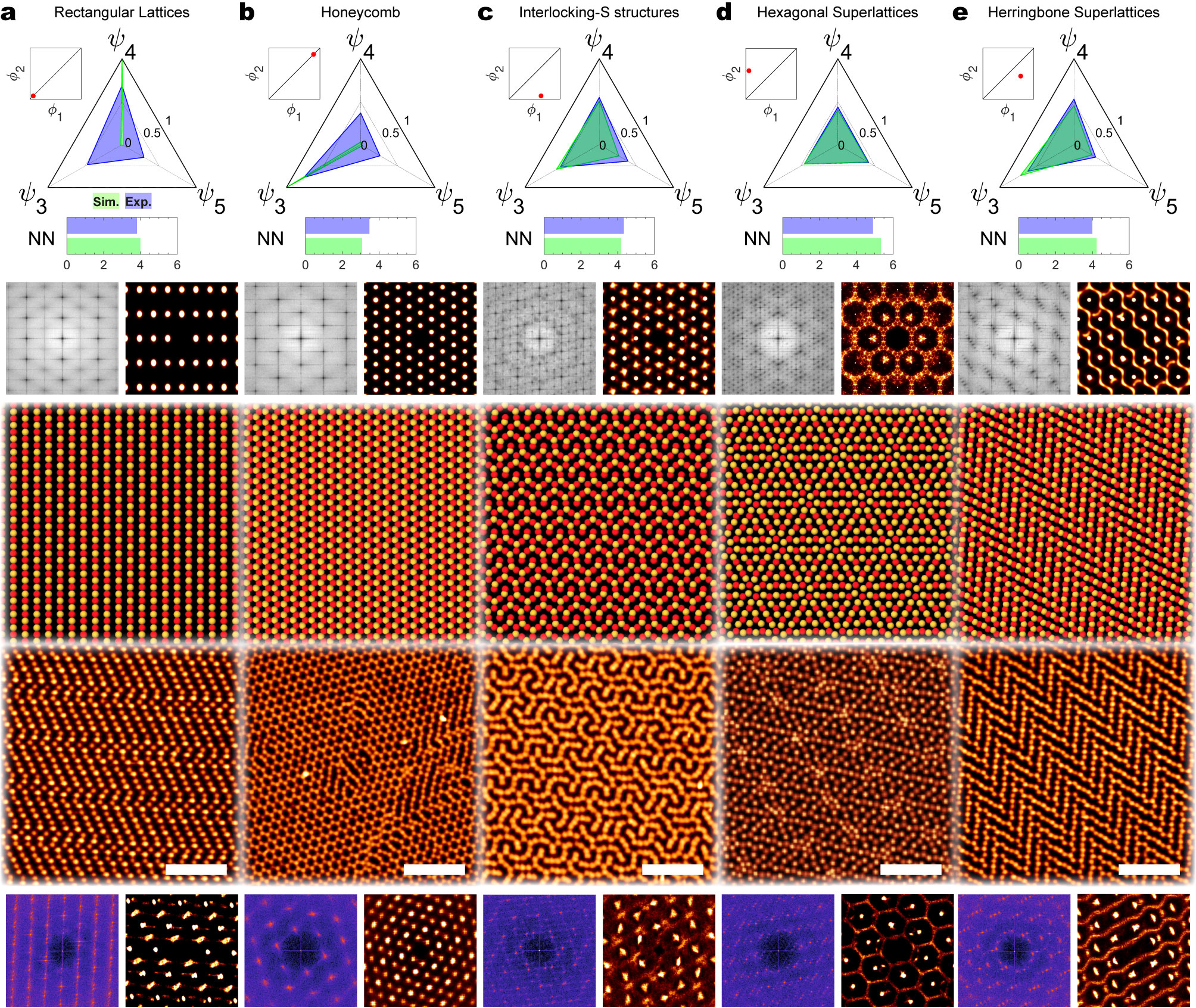}
    \caption{\textbf{Crystalline structures emerging in different regions of the $(\phi_1,\phi_2)$ diagram - experiments (3CS0 microgels) and simulations ($\alpha=1.9$).} \textbf{(a)} Rectangular lattices (0.9,0.9), \textbf{(b)} honeycombs (1.5,1.5), \textbf{(c)} interlocking-S structures (1.35,0.9), and \textbf{(d)} hexagonal (0.9,1.3) and \textbf{(e)} herringbone (1.4,1.2) superlattices. From top to bottom, radar chart of $\psi_3$, $\psi_4$, and $\psi_5$ and bar plot of the number of nearest neighbors (NN) for experimental (blue) and ground-state structures (green); structure factor (left) and 2D positional autocorrelation function (right) of the ground-state structure; simulation snapshot; and AFM image (scale bar 5 $\mu m$) of the corresponding experimental structure accompanied by the structure factor and 2D positional autocorrelation function.}
    \label{fig:Fig3}
\end{figure*}

We investigate the validity of choosing $\alpha<2$ by quantifying the degree and type of crystalline order found in the experimental and simulated structures using the average bond orientational order parameter $\psi_k$ (see Fig. \ref{fig:Fig2} and \ref{fig:Fig3}):
\begin{equation}
  \psi_k=\frac{1}{N}\sum_l^N\left\lvert \frac{1}{N_l}\sum_m^{N_l} e^{ik\theta_{lm}}\right\rvert 
\end{equation}
where $N$ is the total number of particles, $N_l$ is the number of neighbours of particle $l$, $\theta_{l,m}$ is the angle between the unit vector (1,0) and the "bond" vector $r_{lm}$ connecting the reference particle $l$ and its neighbour $m$, and $k$ is the natural number defining the $k$-fold symmetry against which the order parameter is computed. Thus defined, $\psi_k$ is a scalar between 0 and 1 that describes the average degree of $k$-fold symmetry for each particle. For example, a honeycomb lattice corresponds to $\psi_3=1$ because each particle is surrounded by three neighbours that are placed at 120$^\circ$ from each other. Analogously, a square or a rectangular lattice corresponds to $\psi_4=1$. 

We systematically compare experiments with simulations in the parameter space $[0.85,1.65]_{\phi_1,\phi_2} \times [1.4,2.5]_\alpha$ by quantifying the discrepancy between predicted and observed $\psi_k(\phi_1,\phi_2)$ in terms of the overall percentage error $\varepsilon$ (see Methods section). We find that the observed two-dimensional patterns are best described for $\alpha=1.9$ (see Fig. \ref{fig:Fig2}b), in remarkable agreement with the $\alpha=1.8 \pm 0.1$ extracted from the compression experiments of individual monolayers (Fig. \ref{fig:Fig1}d). In particular, not only does $\alpha=1.9$ provide a good quantitative description of our data, but also captures qualitative changes in the type of structures emerging across the ($\phi_1,\phi_2$) diagram (see Figure \ref{fig:Fig3}).

We begin our analysis with patterns for which $\phi_1 = \phi_2$. A key feature captured by the simulations for $\alpha=1.9$ is the transition from rectangular to honeycomb lattices as the total packing fraction ($\phi_1+\phi_2$) is increased while keeping the ratio $\phi_1/\phi_2=1$ (see Fig. \ref{fig:Fig2}c and \ref{fig:Fig3}a-b). By comparing the energy density of honeycomb and rectangular lattices, we find that the latter configuration has lower energy only if $\alpha<2$ and $\phi_1=\phi_2 \lesssim 1$ (see Fig. \ref{fig:Fig2}c). This finding is in good agreement with the observed high values of $\psi_4$ ($>0.6$) in the bottom left corner of the ($\phi_1$,$\phi_2$) diagram and the growth of $\psi_3$ at the expense of $\psi_4$ along the $\phi_1 = \phi_2$ diagonal (see Fig. \ref{fig:Fig2}a). Decreasing $\alpha$ below 1.9 sees further changes in the topology of the phase diagram and, in particular, the emergence of a region in the upper right corner of the ($\phi_1$,$\phi_2$) diagram where $\psi_4>\psi_3$  (see Fig. S4). This region is not observed in the experimental data, hence the existence of an optimal value of $\alpha$.


If combining monolayers with the same packing fraction results in structures with a single type of short-range and long-range order, such as rectangular lattices and honeycombs, mismatching packing fractions ($\phi_1 \neq \phi_2$) bring about a wide range of structures of far greater complexity: superlattices. We identify, in both simulations and experiments, three crystalline structures in different regions of the ($\phi_1,\phi_2$) diagram: interlocking-S structures, hexagonal and herringbone superlattices. These are periodic structures consisting of unit cells spanning several $\sigma$, which present varying symmetries at different length scales, as evident in the real space images, structure factors, and positional autocorrelation functions shown in Figure \ref{fig:Fig3}c-e. 

\begin{figure*}[htbp!]
    \centering
    \includegraphics[width=0.8\textwidth]{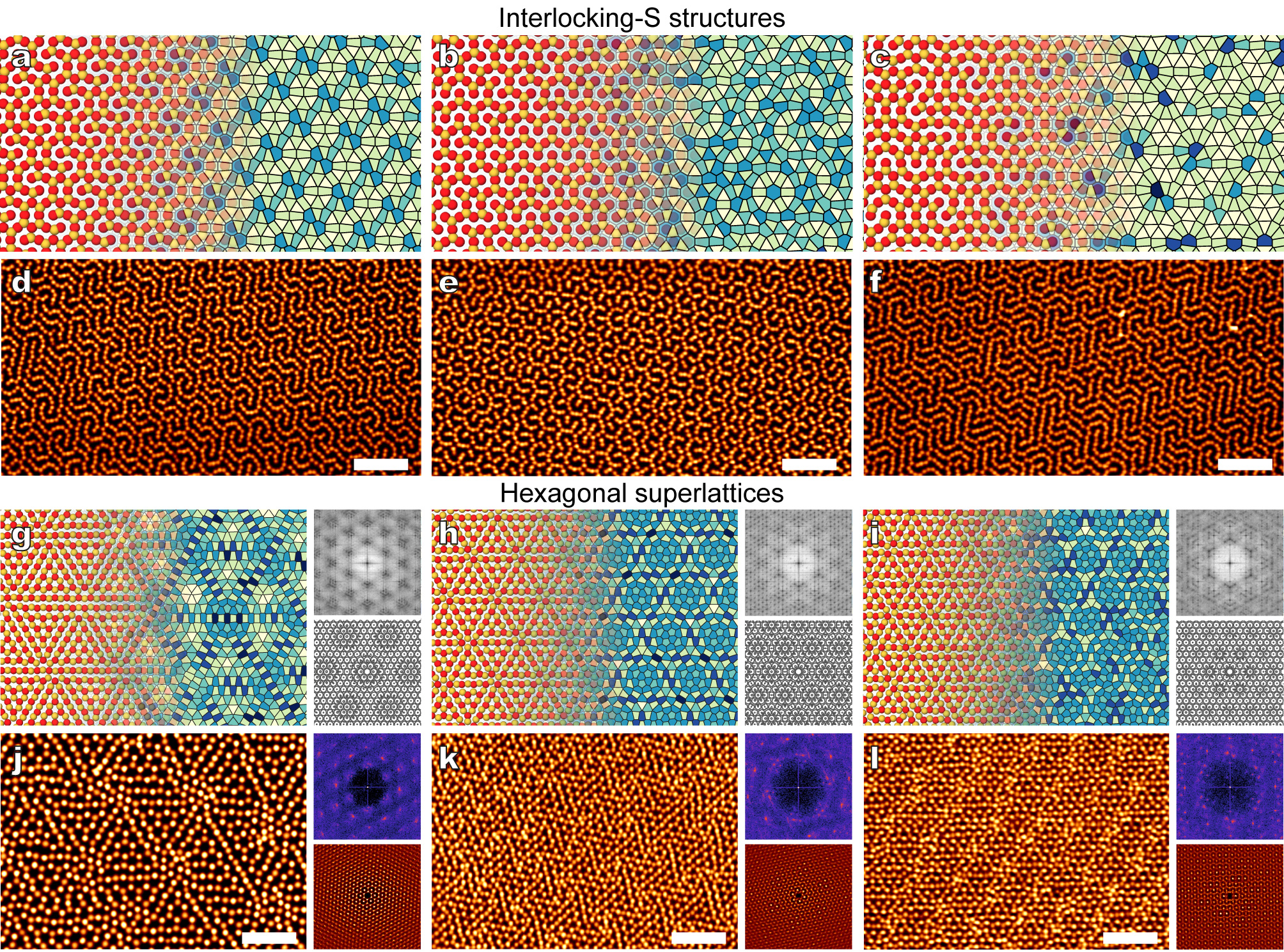}
    \caption{\textbf{Simulated and experimental interlocking-S structures ($\phi_1>\phi_2$) and hexagonal superlattices ($\phi_1<\phi_2$) for different ($\phi_1,\phi_2$) pairs.} \textbf{(a)},\textbf{(b)}, and \textbf{(c)} Snapshots and Voronoi tessellations of ground-state crystalline interlocking-S structures found for $\alpha=1.9$ at (1.4,0.9)$_{\phi_1,\phi_2}$, and their disordered counterparts found at (1.4,0.8)$_{\phi_1,\phi_2}$ and (1.2,0.9)$_{\phi_1,\phi_2}$, respectively. \textbf{(d-f)} AFM images of the corresponding experimental structures obtained with 3CS0 microgels. \textbf{(g-i)} Snapshots, Voronoi tessellations, and structure factor and positional autocorrelation function (top-left and bottom-right of each panel, respectively) of ground-state hexagonal superlattices obtained at a given $\phi_1=0.9$ and varying $\phi_2$, namely 1.1, 1.3, 1.4 from left to right. \textbf{(j-l)} AFM images, structure factor and positional autocorrelation function of the experimental counterparts of \textbf{(g-i)} obtained with 3CS1 (\textbf{g}) and 3CS0 (\textbf{k-l}) microgels. Scale bars: 5 $\mu$m.}
    \label{fig:Fig4}
\end{figure*}

The interlocking-S structures are superlattices occurring at $\phi_1\simeq 1.4 >\phi_2\simeq 0.9$ characterized by a staggered tessellation of rhomboid unit cells (see Figure \ref{fig:Fig4}a), which in turn consist of particles with a coordination number ranging from 3 to 8, encoding tessellations of irregular polygons ranging from triangles to octagons. Perturbing the packing fraction in the neighbourhood of (1.4,0.9)$_{\phi_1,\phi_2}$ disrupts the long-range order, with changes in $\phi_1$ or $\phi_2$ resulting in superstructures of different nature (Figure \ref{fig:Fig4}c-d and Figure \ref{fig:Fig4}e-f). For instance, decreasing $\phi_2$ leads to disordered tessellations of broken dodecagons (the interlocking-S) formed by chains of alternating triangles and rectangles enclosing octagons or hexagons, which locally resemble the Archimedian (3.4.6.4)-tiling also known as the rhombitrihexagonal tiling \cite{grunbaum1989tilings}. On the other hand, decreasing $\phi_1$ results in disordered tessellations reminiscent of the 3-uniform $(3^6; 3^3.4^2; 3.4.6.4)$-tiling, in that they consist of triangles arranged into triangular superstructures bounded by lines of rectangles interrupted by octagons (Figure \ref{fig:Fig4}c and Figure \ref{fig:Fig4}f).

Hexagonal superlattices arise in the same range of $\phi_1+\phi_2$ as the interlocking-S structures, but for $\phi_2>\phi_1$. These are lattices where high-coordinated sites are concentrated in regions that are arranged on a hexagonal lattice formed by triangular superstructures comprising low-coordinated sites. Interestingly, we find that increasing $\phi_2$ translates not only into a higher fraction of high-coordinated sites, particularly the number of pentagons in the Voronoi tessellation, but also into a decrease in the lattice spacing of the hexagonal superlattice (see Figure \ref{fig:Fig4}g-l). 

The herringbone superlattice is instead a simpler structure occurring at high total packing fractions and for $\phi_1\simeq 1.4$ slightly larger than $\phi_2\simeq 1.2$. This structure bears similarities to both honeycombs and rectangular lattices, in that it consists of staggered lines of particles connected by honeycombs. Intermediate but more disordered structures are also found in different regions of the diagram ($\phi_1,\phi_2$), specifically in the neighbourhood of the diagonal (see Fig. S6). 

Finally, we remark that such complex structures are a consequence of the frustration between the second mobile monolayer and the first immobile monolayer, which acts as an effective hexagonal template. In fact, we find that relaxing the constraint on the first monolayer in the molecular dynamics simulations leads to simpler and qualitatively different structures (see  Fig. S8). In the same range of total packing fraction $\phi_1+\phi_2=2.1-2.6$ at which we see the emergence of different superlattices and a wide range of intermediate structures, the ground-state structures of a monolayer of all mobile particles for $\alpha=1.9$ are simple honeycombs.

Our results demonstrate that equilibrium complex patterns can be realized from simple isotropic microscale building blocks by guiding their assembly through sequential steps. In particular, immobilizing a fraction of the particles converts them into a template that regulates the assembly of the rest of the population. This process enables both the overcoming of kinetic trapping and the realization of new structures stemming from the frustration between the arrangements of both populations. Kinetic limitations can be overcome by very small nanoparticles, where thermal agitation constantly equilibrates the system during controlled assembly protocols \cite{Murray_AnnRev_2000,Talapin_2009}, but they pose severe hurdles for (sub)micron colloids\cite{Leunissen_2005}. Two-dimensional complex patterns with structural motifs over these length scales are for instance sought to realize biomimetic surfaces \cite{Kraus_biomimetic_assembly_2013} or metasurfaces with emerging optical \cite{Mayer_plsamonic_metasurfaces_2019} and mechanical properties \cite{Jinwoong_2018}, where fine structural control over large areas is required. 

Our findings reveal that this degree of complexity and control can be achieved through the rational design of short-range soft repulsive potentials and self-templating protocols. Uniform target structures can be deposited over large areas by keeping the surface pressure constant throughout the deposition process (Fig. S9). Even though the detailed effect of potential and force concavity as soft particles overlap has received deep theoretical attention \cite{Likos_PhysRevE_2001}, its exploitation for the synthesis of colloids with tailored interactions upon two-dimensional confinement, e.g. at a fluid interface, remains largely untapped. As an example, numerical simulations indicate that colloids coated by linear amphiphilic polymer shells that are adsorbed at a water-oil interface interact via Gaussian potentials \cite{Schwenke_2014}, but controlling the interplay between chemistry and cross-linking density profiles for microgels confined at interfaces to design pair potentials is an open field. For instance, we hypothesize that different microgel architectures can be conceived so as to realize particles interacting via GHP with tuneable $\alpha$. Finally, self-templating and sequential depositions allow for the integration of soft building blocks with varying softness \cite{Rey_JACS_2017}, size \cite{Fernandez-Rodriguez_Nanosacle_2018} and material composition \cite{Honold_2016}, paving the way towards the robust and versatile fabrication of functional two-dimensional patterns.


\section*{Acknowledgements}
The authors thank Emanuela Zaccarelli, Fabrizio Camerin, Walter Steurer and Thomas Weber for discussions. LI and MAFR acknowledge financial support from the Swiss National Science Foundation Grant PP00P2-172913/1.

\section*{Author Contributions}
Author contributions are defined based on the CRediT (Contributor Roles Taxonomy) and listed alphabetically. Conceptualization: MNA, MAFR, FG, LI. Formal analysis: MNA, MAFR, FG. Funding acquisition: LI. Investigation: MNA, MAFR, DG, FG. Methodology: MNA, MAFR, FG, LI. Project administration: LI. Software: MAFR, FG. Supervision: MAFR, LI. Validation: MNA, MAFR, DG, FG. Visualization: MNA, MAFR, FG, LI. Writing – original draft: MAFR, FG, LI. Writing – review and editing: MNA, MAFR, DG, FG, LI. 

\section*{Supporting Information}
Supporting Information is available for this paper. Correspondence and requests for materials should be addressed to fabio.grillo@mat.ethz.ch, ma.fernandez@mat.ethz.ch and lucio.isa@mat.ethz.ch. 

\section*{Data availability statement}
The data that support the findings of this study are available from the corresponding author upon reasonable request.

\section*{Code availability statement
}
Numerical simulations and analysis code that support the findings of this study are available from the corresponding author upon reasonable request.

\section*{Methods}
\subsection*{Microgel synthesis}
We synthesized PNIPAM microgels via aqueous one-pot precipitation polymerization approach\cite{mgel_synthesis}. Two amounts of the crosslinker N-N’-Methylenebisacrylamide (BIS) were added to 180 mM N-isopropylacrylamide (NiPAm) monomer solutions to obtain microgels with crosslinker-to-monomer ratios of 3 wt$\%$ and 5 wt$\%$, respectively. NiPAm and BIS were dissolved in MilliQ water at 80 $^\circ$C wit the aid of magnetic stirring. The solution was then deaerated with N2, before the addition of 1.8 mM potassium persulfate (KPS) to initiate the reaction. The temperature was maintained at 80 $^\circ$C for 5 h to ensure that it ran to completion. The resultant microgels were then cleaned with three ultracentrifugation cycles at 20000 rpm for 1 h. At the end of each ultracentrifugation cycle, the supernatant was removed and replaced with fresh MilliQ water, and the microgels were re-dispersed by 1 h of ultrasonication. Microgels obtained via this synthesis process present a core-shell morphology, with a degree of crosslinking that decreases radially \cite{mgel_shape}. Additional particle types were synthesized by further extending the PNiPam shells of the microgels via an extra growth step. NiPAm and BIS in the corresponding crosslinker-to- mass ratios of 3 wt$\%$ and 5 wt$\%$ were dissolved in 6 mL of MilliQ water under N$_2$ flow with the aid of magnetic stirring. In a separated vessel, 0.25 of the previously synthesized, freeze-dried microgels were re-dispersed in 30 mL of MilliQ water under a flow of $N_2$ and heated to 80 $^\circ$C under magnetic stirring. The reaction was performed in four steps within the vessel containing the dispersed microgels, by adding 1 mL of 1.2 mM KPS in MilliQ water and 1.25 mL of the crosslinker-monomer solution every 10 minutes, for a total of 40 minutes. Afterwards, the reaction mixture was kept at 80 $^\circ$C under magnetic stirring for 5 h. By following this protocol, we could vary the microgel size at a given crosslinking mass ratio. In particular, we label our particles as CXSY, where X = 3 or 5 is the crosslinking mass ratio and Y=0 or 1 is the number of steps for additional shell growth. The size of the microgels in bulk MilliQ-water was measured by dynamic light scattering (DLS, Malvern Zetasizer) at 25 $^\circ$C (see Table \ref{table:mgel_size}).

\subsection*{Deposition of monolayers from liquid-liquid interfaces}
Monolayers of microgels were deposited onto 2x2 cm$^2$ silicon substrates (Siltronix, <100> 100 mm single polished side) following the procedure described in a previous work \cite{Rey_SoftMatter_2016}. In brief, we prepared 0.1 wt\% microgel suspensions in 4:1 water:isopropanol mixtures. The presence of isopropanol assists the spreading of microgels at the fluid interface. Prior to deposition, the silicon substrates were rinsed in three consecutive ultrasonic baths of toluene, isopropanol and MilliQ water and then dried with pressurized N$_2$. We positioned the silicon substrates inside a customized liquid-liquid Langmuir-Blodgett trough (KSV5000, Biolin Scientific), by connecting them to the dipping arm at an angle of 30$^\circ$ relative to the interface plane. We filled the trough with MilliQ water until the substrate was fully immersed and the water reached the position of the barriers. Next, 100 mL of n-hexane were added to create the water/hexane interface, we raised the substrate until it just crossed the water/hexane interface and zeroed the surface pressure $\Pi$. The point where the wafer intersected the interface was used as a reference to reconstruct the value of surface pressure as a function of position.  We then added the desired amount of the microgel dispersion to the interface with a Hamilton glass microsyringe (100 $\mu L$) and let the system stabilize for 10 minutes. Finally, $\Pi$ was gradually increased by compressing the interface with the barriers from 197.5 to 59.5 cm$^2$ at a rate of 2.3 mm/min, while the dipping arm was raised at 0.5 mm/min. The second deposition was repeated with the same protocol by using the same substrate with the previously deposited monolayer but rotating it by 90$^\circ$ with respect to the direction of the first compression. In this way, we created two bands where only a single monolayer was present (Fig \ref{fig:Fig1}b), which can be used for the estimation of the surface densities of the first $\phi_1$ and second $\phi_2$ depositions, as well as the total density $\phi_1$+$\phi_2$. The deposition of target structures over large areas (see Fig. S9) was achieved by keeping the surface pressure constant throughout the deposition process via the feedback control loop of the Langmuir-Blodgett. 

The microstructure of the monolayers was imaged via atomic force microscopy (AFM, Brucker Icon Dimension) in tapping mode (cantilevers resonance frequency: 300 kHz, spring constant: 26  mN/m). AFM images of 40x40 and 88x88 $\mu m^2$ were taken at a rate of 1 Hz. Compression curves were constructed by relating the position-dependent $\Pi$ and area per particle $A_p$, obtained by extracting the number of particles in each image with ImageJ, and converted to $\Pi$ vs $r/\sigma$ (Fig \ref{fig:Fig1}d), where $r$ is the center-to-center distance between microgels, assuming hexagonal packing. The diameter of the microgels at the interface measured from isolated deposited microgels by limiting the height in the AFM images to 5 nm to increase the contrast of the thinner corona is shown in Table \ref{table:mgel_size}.

\begin{table}[]
    \centering
    \begin{tabular}{|c|c|c|c|}\hline
             & $\sigma_{DLS}\,(nm)$ & $\sigma_{AFM}\,(nm)$ & $\sigma_{fit}\,(\mu m)$ \\\hline
    3CS0     & $618\pm83$ & $923\pm64$ & $1.11$ \\\hline
    3CS1     & $879\pm121$ & $1578\pm46$ & $1.52$ \\\hline
    5CS1     & $620\pm204$ & $1066\pm71$ & $1.10$ \\\hline
    \end{tabular}
    \caption{Diameter $\sigma$ of the microgels in bulk (DLS), at the interface (from 25 isolated deposited microgels, by AFM), and fitted from the compression curves.}
    \label{table:mgel_size}
\end{table}

\subsection*{Reagents}
N-isopropylacrylamide (NIPAM, TCI 98.0\%), N-N’-Methylenebisacrylamide (BIS, Fluka 99.0\%), potassium persulfate (KPS, Sigma-Aldrich 99.0\%), isopropanol (Fisher Chemical, 99.97\%), toluene (Fluka Analytical, 99.7\%) and n-hexane (Sigma-Aldrich, HPLC grade 95\%). The monomer was purified by recrystallisation, in 60/40 v/v toluene/hexane.  The rest of the reagents were used without further purification.

\subsection*{Simulations}
The ground-state structures were estimated through two-dimensional molecular dynamics simulations carried out with the open source libraries of the simulation toolkit HOOMD-blue~\cite{RN3697,RN3698}. We ran simulations for 10x10 evenly spaced ($_{\phi_1}$,${\phi_2}$) pairs in the interval [0.85,1.6]$_{\phi_1}$ x[0.85,1.65]$_{\phi_2}$ for each of the following values of $\alpha$: 1.4, 1.5, 1.6, 1.7, 1.8, 1.9, 2, 2.25, 2.5. The equations of motion were integrated in the canonical NVT ensemble using the Nosé-Hoover thermostat with a time step dt=0.01 and a coupling constant $\tau=1$, both expressed in terms of normalized time units $\sqrt{M\sigma^2/\epsilon}=1$, where $M$ and $\sigma$ are the mass and diameter of the particles, and $\epsilon$ is the energy scale of the interaction potential, which, without loss of generality, were all set equal to 1. For each ($\phi_1$,$\phi_2$) pair, the simulations were initialized by placing two populations of equally-sized spherical particles in a 2D box: one of immobile particles constrained to a hexagonal lattice, and the other of randomly distributed mobile particles. Periodic boundary conditions were enforced on all the sides of the box. The total number of particles varied in the range 6238-11250. The position of the mobile particles was let evolve towards the minimum energy configuration by varying their reduced temperature $kT/\epsilon$ from $10^{-2}$ to 0 over a period of 1.2x10$^8$ steps. In particular, the ground-state configurations were obtained via the Fast Inertial Relaxation Engine (FIRE) algorithm~\cite{RN3696}, which was used to bring $kT/\epsilon$ from $10^{-4}$ to 0 while minimizing the total potential energy. At all times, the mobile particles interacted with each other and with the immobile particles according to the same interaction potential, namely the generalized Hertzian potential defined in Equation \ref{eq:GHP}.

To minimize the formation of meta-stable grain-boundaries and test the reproducibility of the ground-state configurations, we carried out two consecutive annealing cycles: $kT/\epsilon$ is first brought to zero from $10^{-2}$ over a period of $2x10^7$, then brought back to $0.5x10^{-2}$, and eventually decreased to 0 after a long period of $1x10^8$ steps. As shown in Fig. S7, the bond orientational order parameters $\phi_k$ attain the same values at the end of both annealing cycles, attesting to reproducibility of the observed symmetries.

\subsection*{Structural analysis}
The bond orientational order parameters $\psi_k$, the positional correlation function, and the Voronoi tessellations were computed using several modules of the freud library~\cite{eric_s_harper_2016_166564}. The structure factors were constructed by calculating the fast Fourier transform of the particles' positions. Because the observed structures present more than one type of symmetry, and thus different characteristic length scales, the nearest neighbours of each particle were identified based on the Voronoi tessellation rather than on a single cut-off distance. The nearest neighbours of the $i$-th particle are defined as the $j$-particles whose Voronoi cell share an edge with the cell of the $i$-th particle. Only edges greater than 8\% of the perimeter of the $i$-th cell are considered. This threshold was introduced to reduce the sensitivity of the computation to small lattice distortions, and thus to consider only the most representative bonds. 

The values of $\psi_k$ for 3CS0 and 3CS1 were estimated based on the particle's positions extracted from 250-350 AFM 40x40 $\mu$m$^2$ images per sample. The corresponding values of $\phi_1$ and $\phi_2$ were extrapolated from AFM images taken in the corresponding lateral bands of the wafer where only one monolayer was deposited. To test the validity of this approximation we measured the ($\phi_1$,$\phi_2$) pairs of 40-50 AFM images where the particles of the first and the second deposition could be singled out based on a slight height difference between the two populations. As shown in Fig. S3, the actual values of $\phi_1$ and $\phi_2$ are in good agreement with the extrapolated ones, with a root-mean-square error (RMSE) of $\simeq 0.05$. The latter was used as a measure of the uncertainty in $\phi_1$ and $\phi_2$. Assuming the true values of the latter to follow a normal distribution centered in the extrapolated values with standard deviation equal to the RMSE, the 95\% confidence intervals were estimated to be equal to the extrapolated $\phi_i\pm 0.1$. 

To reconstruct the best estimate of $\psi_k(\phi_1,\phi_2)$ we propagated the uncertainty in $\phi_1$ and $\phi_2$ via the Monte Carlo method by averaging 10000 realizations of $\psi_k(\phi_1+\xi,\phi_2+\eta)$, where $\xi$ and $\eta$ are normally distributed random numbers with zero mean and a standard deviation of 0.05. In particular, all the realizations of $\psi_k$ were interpolated and averaged at 100x100 evenly spaced points in the interval [0.85,1.65]$_{\phi_1}$ x[0.85,1.65]$_{\phi_2}$ via thin plate splines. The values of $\psi_k$ thus estimated were then compared with the ones obtained for the simulated ground-state structures, which were also interpolated at the same ($\phi_1$,$\phi_2$) pairs. 

The agreement between simulated and experimental structures was quantified in terms of the median symmetric accuracy\cite{Morley2018}: $\zeta_k=100\left(\exp{(M\left\lvert \ln{(\psi_k^{sim}/\psi_k^{exp})}\right\rvert}-1\right)$, where $\psi_k^{sim}$ and  $\psi_k^{exp}$ are the predicted and observed values of $\psi_k$, respectively, and $M$ is the median function. This metric, which can be interpreted as a percentage error, is insensitive to outliers and gives the same weight to overprediction and underprediction, and thus provides a robust and unbiased measure of the accuracy of our predictions. The overall percentage error $\varepsilon$ is then defined as the weighted average of $\zeta_k$: $\varepsilon=\sum_{k=3}^8 w_k\zeta_k$, with weights proportional to the interquantile range (IQR) of the respective observed $\psi_k$: $w_k=\mathrm{IQR}_k/\sum_{k=3}^8 \mathrm{IQR}_k$. This is to give more weight to the most representative observables, that is the ones that vary the most across ($\phi_1$,$\phi_2$) pairs. In fact, the experimental $\psi_k$ attain the highest values and degree of variation for $k=3$ and $k=4$ across all ($\phi_1$,$\phi_2$) pairs, and rapidly drops for $k>4$, approaching a virtually constant value between 0.15-0.2 for $k>8$ (see Fig. S10). 

\bibliography{microgels}
\end{document}